\documentclass[twocolumn,preprintnumbers,amsmath,amssymbm,prd]{revtex4}
\usepackage{epsfig}
\usepackage{graphicx}

\begin{document}
\title{The instability spectrum of weakly-magnetized SU(2) Reissner-Nordstr\"om black holes}
\author{Shahar Hod}
\address{The Ruppin Academic Center, Emeq Hefer 40250, Israel}
\address{ }
\address{The Hadassah Institute, Jerusalem 91010, Israel}
\date{\today}

\begin{abstract}
\ \ \ It is well known that the U(1) Reissner-Nordstr\"om black hole
is stable within the framework of the Einstein-Maxwell theory.
However, the SU(2) Reissner-Nordstr\"om black-hole solution of the
coupled Einstein-Yang-Mills equations is known to be unstable. In
fact, this magnetically charged black hole is characterized by an
infinite set of unstable (growing in time) perturbation modes. In
the present paper we study {\it analytically} the instability
resonance spectrum of weakly-magnetized SU(2) Reissner-Nordstr\"om
black holes. In particular, we obtain explicit analytical
expressions for the infinite set $\{\omega_n\}_{n=0}^{n=\infty}$ of
imaginary eigenvalues that characterize the instability growth rates
of the perturbation modes. We discuss the role played by these
unstable eigenvalues as critical exponents in the gravitational
collapse of the Yang-Mills field. Finally, it is shown that our
analytical formulas for the characteristic black-hole instability
spectrum agree with new numerical data that recently appeared in the
literature.
\end{abstract}
\bigskip
\maketitle


\section{Introduction}

The Einstein-Yang-Mills theory has attracted much attention from
both physicists and mathematicians since the discovery, made by
Bartnik and McKinnon \cite{BarMck}, of a discrete family of regular
self-gravitating solitonic solutions of the coupled equations. As
shown by Bizo\'n \cite{Biz} (see also \cite{Gal}), the theory also
admits a discrete family of hairy black-hole solutions, known as
colored black holes \cite{Notehn}. In fact, it was shown by Yasskin
\cite{Yas} already in the 70s that the coupled Einstein-Yang-Mills
equations admit an explicit solution in the form of a magnetically
charged Reissner-Nordstr\"om black hole.

These solutions of the coupled Einstein-Yang-Mills equations are
known to be {\it unstable} \cite{Str,BizWal,Mas}. In particular, the
$n$th colored black-hole solution
is characterized by $n$ unstable (growing in time) perturbation
modes \cite{Notepmm}. As for the magnetized SU(2)
Reissner-Nordstr\"om black hole, it was proved in \cite{Mas} that
this unstable solution of the coupled Einstein-Yang-Mills equations
is characterized by an {\it infinite} set of unstable perturbation
modes. This fact is quite surprising since the U(1)
Reissner-Nordstr\"om black hole is known to be stable within the
framework of the coupled Einstein-Maxwell theory \cite{Mon} (see
also \cite{Hods}).

So why is it interesting to study these unstable solutions of the
Einstein-Yang-Mills theory? One important reason lies in the fact
that these unstable configurations have been identified as {\it
critical} solutions \cite{GunMar} of the coupled Einstein-Yang-Mills
equations \cite{ChCh,ChHi,BizCha,Oliv}. That is, these unstable
configurations play the role of intermediate attractors in the
dynamical gravitational collapse of the Yang-Mills field
\cite{Noteot}.

In particular, it has been demonstrated numerically
\cite{ChCh,ChHi,BizCha,Oliv} that, during a near-critical
gravitational collapse of the Yang-Mills field, the time spent in
the vicinity of the critical solution (that is, the time spent in
the vicinity of an unstable black-hole configuration of the
Einstein-Yang-Mills theory) exhibits a critical scaling behavior of
the form \cite{Notepp}
\begin{equation}\label{Eq1}
\tau=\text{const}-\gamma\ln|p-p^*|\  ,
\end{equation}
where the critical exponents are directly related to the instability
eigenvalues that characterize the relevant unstable black-hole
(critical) solution \cite{ChCh,ChHi,BizCha,Oliv}:
\begin{equation}\label{Eq2}
\gamma=1/\omega_{\text{instability}}\  .
\end{equation}

It is therefore of physical interest to study the instability
spectra which characterize the black-hole solutions of the coupled
Einstein-Yang-Mills equations. A detailed numerical study of the
instability spectrum of the $n=1$ colored black holes can be found
in \cite{BizCha}. Most recently, Rinne \cite{Oliv} has computed
numerically the instability eigenvalues which characterize the SU(2)
Reissner-Nordstr\"om black holes in the framework of the
Einstein-Yang-Mills theory \cite{NoteRin}.

The main goal of the present paper is to determine {\it
analytically} the instability spectrum (that is, the {\it infinite}
set of imaginary eigenvalues) which characterizes the SU(2)
Reissner-Nordstr\"om black-hole spacetime. As we shall show below,
the Schr\"odinger-like wave equation [see Eq. (\ref{Eq6}) below]
which governs the dynamics of linear perturbations to the SU(2)
Reissner-Nordstr\"om black-hole spacetime is amenable to an
analytical treatment in the regime of weakly-magnetized black holes
\cite{Notelms}.

\section{Description of the system}

The SU(2) Reissner-Nordstr\"om black-hole solution with unit
magnetic charge is described by the line element \cite{Yas}
\begin{equation}\label{Eq3}
ds^2=-\Big(1-{{2m}\over{r}}\Big)dt^2+\Big(1-{{2m}\over{r}}\Big)^{-1}dr^2+r^2(d\theta^2+\sin^2\theta
d\phi^2)\ ,
\end{equation}
where the mass function $m=m(r)$ is given by \cite{Noteunit}
\begin{equation}\label{Eq4}
m(r)=M-{{1}\over{2r}}\  .
\end{equation}
The black-hole outer horizon is located at
\begin{equation}\label{Eq5}
r_+=M+\sqrt{M^2-1}\  .
\end{equation}

The dynamics of linearized perturbations $\xi(r) e^{-i\omega t}$ of
the black-hole spacetime is governed by the Schr\"odinger-like wave
equation \cite{Bizw}
\begin{equation}\label{Eq6}
\Big[{{d^2}\over{dx^2}}+\omega^2-U(x) \Big]\xi=0\  ,
\end{equation}
where the ``tortoise" radial coordinate $x$ is defined by
\cite{Notehor}
\begin{equation}\label{Eq7}
dx/dr=[1-2m(r)/r]^{-1}\  ,
\end{equation}
and the effective binding potential in (\ref{Eq6}) is given by
\begin{equation}\label{Eq8}
U[x(r)]=-{1 \over {r^2}}\Big[1-{{2m(r)} \over r}\Big] \  .
\end{equation}
Note that unstable (growing in time) modes are characterized by
\begin{equation}\label{Eq9}
\Im\omega>0\  .
\end{equation}
These unstable modes may equivalently be regarded as `bound states'
(characterized by $\omega^2<0$) of the effective binding potential
(\ref{Eq8}).

\section{The large-mass limit}

In this paper we shall consider weakly-magnetized black holes whose
unit magnetic charge is small on the scale set by the black-hole
mass:
\begin{equation}\label{Eq10}
M\gg1\  .
\end{equation}
The Schr\"odinger-like wave equation (\ref{Eq6}) for these large
(weakly-magnetized) black holes can be approximated by
\begin{equation}\label{Eq11}
\Big[{{d^2}\over{dx^2}}+\omega^2+\Big(1-{{2M}\over{r}}\Big){{1}\over{r^2}}
\Big]\xi=0\ .
\end{equation}

The effective potential in (\ref{Eq11}) is negative in the entire
range $-\infty<x<\infty$ and it vanishes asymptotically for
$x\to\pm\infty$. As noted in \cite{Bizw}, this fact guarantees that
the Schr\"odinger-like wave equation (\ref{Eq11}) possesses at least
one {\it unstable} mode with $\omega^2<0$ (that is, at least one
bound state with negative energy \cite{Notewe}). In fact, as we
shall show below, the Schr\"odinger-like wave equation (\ref{Eq11})
is characterized by an {\it infinite} set
$\{\omega_n\}_{n=0}^{n=\infty}$ of unstable modes (an infinite set
of bound-state resonances) with $\Im\omega>0$.

As we shall now show, the wave equation (\ref{Eq11}) is amenable to
an analytical treatment in the regime (\ref{Eq10}) of large
black-hole masses (that is, in the regime of weakly-magnetized black
holes). We first point out that the Schr\"odinger-like equation
(\ref{Eq11}) is of the same form as the familiar Regge-Wheeler
equation \cite{Whee}
\begin{equation}\label{Eq12}
\Big[{{d^2}\over{dx^2}}+\omega^2-\Big(1-{{2M}\over{r}}\Big){{l(l+1)}\over{r^2}}
\Big]\xi=0\
\end{equation}
which describes electromagnetic perturbations of frequency $\omega$
and angular harmonic index $l$ in the Schwarzschild black-hole
spacetime. In our case, the effective harmonic index acquires a
complex value [compare Eqs. (\ref{Eq11}) and (\ref{Eq12})]
\cite{Notepm}:
\begin{equation}\label{Eq13}
\ell\equiv l_{\text{eff}}={{-1+i\sqrt{3}}\over{2}}
\  .
\end{equation}

\section{The fundamental instability eigenvalue}

In order to calculate the fundamental instability eigenvalue
$\omega_0$ of the system, we shall closely follow the analysis of
Dolan and Ottewill \cite{Dol} who provided an elegant method for the
calculation of the fundamental black-hole {\it quasinormal
frequencies}. In this section we shall demonstrate that this
analytical method can also be applied successfully to the analysis
of fundamental {\it bound-state energies} (in our case, for the
calculation of the fundamental instability eigenvalue)
\cite{Notebc}.

The analytical approach of \cite{Dol} is based on an expansion of
the resonances in inverse powers of the harmonic parameter $L\equiv
l+1/2$:
\begin{equation}\label{Eq14}
M\omega_{ln}=\sum_{k=-1}^{\infty}w_{k}L^{-k}\  ,
\end{equation}
where the expansion coefficients $\{w_{k}\}_{k=-1}^{k=4}$ are given
by equations (17)-(22) of \cite{Dol}. Substituting
\begin{equation}\label{Eq15}
L=\ell+{1\over 2}=i{{\sqrt{3}}\over{2}}
\end{equation}
into (\ref{Eq14}), one finds \cite{Noteln}
\begin{equation}\label{Eq16}
M\omega^{\text{ana}}_0=i\times
{{413545392-108984521\sqrt{3}}\over{1836660096}}\simeq 0.1224i
\
\end{equation}
for the fundamental \cite{Notefun} instability eigenvalue of the
black hole.

For comparison, the numerically computed fundamental eigenvalue in
the large-mass limit (\ref{Eq10}) is given by \cite{Oliv,Notelarg}
\begin{equation}\label{Eq17}
M\omega^{\text{num}}_0=0.1243i
\  .
\end{equation}
One therefore finds a fairly good agreement (to better than $2\%$),
\begin{equation}\label{Eq18}
{{\omega^{\text{ana}}_0}\over{\omega^{\text{num}}_0}}\simeq 0.985\ ,
\end{equation}
between the {\it analytically} calculated fundamental eigenvalue
(\ref{Eq16}) and the corresponding {\it numerically} computed
\cite{Oliv} eigenvalue (\ref{Eq17}).

As discussed in \cite{Dol}, the validity of the expansion method
(\ref{Eq14}) is restricted to the fundamental $n\lesssim l$ modes.
In the next section we shall develop a different analytical approach
in order to explore the rest of the (infinitely large) family of
unstable resonances $\{\omega_n\}_{n=1}^{n=\infty}$ \cite{Noteex}.

\section{The infinite spectrum of unstable bound-state resonances}

Taking cognizance of (\ref{Eq16}), one realizes that the entire
instability spectrum of the weakly-magnetized black holes is
characterized by the relation $M|\omega|<1$ \cite{Notesat}. In fact,
the numerical results of Rinne \cite{Oliv} indicate that the excited
eigenvalues $\{\omega_n\}_{n=1}^{n=\infty}$ of the system are
characterized by the property
\begin{equation}\label{Eq19}
M|\omega_n|\ll1\ \ \ ; \ \ \ n=1,2,3,...
\end{equation}
As we shall now show, a {\it low-frequency} analysis of the
perturbation modes can yield, with a remarkably good accuracy, the
excited eigenvalues $\{\omega_n\}_{n=1}^{n=\infty}$ of the unstable
black hole.

As shown in \cite{Fab}, the Regge-Wheeler equation (\ref{Eq12}) is
amenable to an analytical treatment in the low-frequency regime
(\ref{Eq19}). In particular, the absorption and reflection
coefficients of scattered low-frequency electromagnetic waves in a
spherically-symmetric black-hole spacetime were calculated in
\cite{Fab}. The analytical method used in \cite{Fab} can be
summarized as follows: (1) find approximate solutions of Eq.
(\ref{Eq12}) in three spatially distinct regions of the black-hole
exterior region, and then (2) use {\it two} matching procedures
(which are based on continuity conditions) in order to overlap the
three analytical solutions (see \cite{Fab} for details).

While the analytical technique used in \cite{Fab} for the analysis
of the low-frequency {\it scattering} problem can also be applied to
the analysis of the {\it bound-state} resonances of Eq.
(\ref{Eq11}), here we shall use a somewhat simpler trick which
involves a {\it single} (rather than a {\it double} \cite{Fab})
matching procedure.

The trick is to analyze the {\it physically equivalent} Teukolsky
radial equation \cite{TeuPre}:
\begin{equation}\label{Eq20}
\Delta^2{{d^2\psi}\over{dr^2}}+\Big[{{\omega^2r^4+2iM\omega
r^2}}-\Delta[2i\omega r+\ell(\ell+1)]\Big]\psi=0\ ,
\end{equation}
which, like Eq. (\ref{Eq12}), describes the dynamics of
electromagnetic perturbation fields in the non-rotating black-hole
spacetime \cite{Noteas}. Here $\Delta\equiv r^2-2Mr$ and in our case
$\ell$ is given by Eq. (\ref{Eq13}). It was first proved by
Chandrasekhar \cite{Chan} that the Teukolsky radial equation
(\ref{Eq20}) for non-rotating black holes (also known as the
Bardeen-Press equation \cite{BarPre}) is physically equivalent
to the Regge-Wheeler equation (\ref{Eq12}).

As we shall now show, one can derive analytically the entire low
frequency instability spectrum $\{\omega_n\}_{n=1}^{n=\infty}$ of
the black hole from Eq. (\ref{Eq20}) using a single matching
procedure [instead of the double matching procedure required for the
analysis of Eq. (\ref{Eq12})] \cite{Notesin,Hodcen}. We shall look
for
bound-state ($\omega^2<0$) solutions which are characterized by
\begin{equation}\label{Eq21}
\psi(x\to -\infty)\sim e^{|\omega|x}\to 0\  ,
\end{equation}
and
\begin{equation}\label{Eq22}
\psi(x\to\infty)\sim xe^{-|\omega|x}\to 0\  ,
\end{equation}
where $\omega=i|\omega|$.

It proves useful to define new dimensionless variables
\cite{Hodcen,Page}
\begin{equation}\label{Eq23}
z\equiv {{r-2M}\over {2M}}\ \ \ ; \ \ \ k\equiv -2iM\omega\ ,
\end{equation}
in terms of which the wave equation (\ref{Eq20}) becomes
\begin{eqnarray}\label{Eq24}
z^2(z+1)^2{{d^2\psi}\over{dz^2}}\nonumber \\
+\big[-k^2z^4+2kz^3-\ell(\ell+1)z(z+1)-k(2z+1)-k^2\big]\psi&=&0\ . \nonumber \\
\end{eqnarray}

The solution of the radial equation (\ref{Eq24}) in the near-horizon
region $kz\ll 1$ which satisfies the boundary condition (\ref{Eq21})
is given by \cite{Hodcen,Page}
\begin{eqnarray}\label{Eq25}
\psi(z)=z^{1+k}(z+1)^{1-k} {_2F_1}(-\ell+1,\ell+2;2+2k;-z)\  , \nonumber \\
\end{eqnarray}
where $_2F_1(a,b;c;z)$ is the hypergeometric function \cite{Abram}.

The solution of the radial equation (\ref{Eq24}) in the far-region
$z\gg1$ is given by \cite{Hodcen,Page}
\begin{eqnarray}\label{Eq26}
\psi(z)=Ae^{kz}z^{\ell+1}{_1F_1}(\ell+2;2\ell+2;-2kz)\nonumber
\\+Be^{kz}z^{-\ell}{_1F_1}(-\ell+1;-2\ell;-2kz)\ ,
\end{eqnarray}
where $_1F_1(a;c;z)$ is the confluent hypergeometric function
\cite{Abram} and the coefficients $\{A,B\}$ are constants. These
coefficients can be determined by matching the two solutions for the
radial function, (\ref{Eq25}) and (\ref{Eq26}), in the overlap
region \cite{Noteov}
\begin{equation}\label{Eq27}
1\ll z\ll 1/k\  .
\end{equation}
This matching procedure yields \cite{Hodcen,Page}
\begin{equation}\label{Eq28}
A={{\Gamma(2\ell+1)\Gamma(2+2k)}\over
{\Gamma(\ell+2)\Gamma(\ell+1+2k)}}\
 ,
\end{equation}
and
\begin{equation}\label{Eq29}
B={{\Gamma(-2\ell-1)\Gamma(2+2k)}\over
{\Gamma(-\ell+1)\Gamma(-\ell+2k)}}\ .
\end{equation}

Finally, substituting (\ref{Eq28}) and (\ref{Eq29}) into the
far-region solution (\ref{Eq26}) and using the asymptotic ($z\gg1$)
form of the confluent hypergeometric functions \cite{Abram}, one
finds \cite{Hodcen,Page}
\begin{eqnarray}\label{Eq30}
\psi(z\to\infty)=\psi_1 re^{-kz}+\psi_2 r^{-1}e^{kz}\  ,
\end{eqnarray}
where
\begin{eqnarray}\label{Eq31}
\psi_1={{(2\ell+1)\Gamma^2(2\ell+1)\Gamma(2+2k)}\over{2\Gamma^2(\ell+2)
\Gamma(\ell+1+2k)}}(-2k)^{-\ell}M^{-1} \nonumber
\\
-{{(2\ell+1)\Gamma^2(-2\ell-1)\Gamma(2+2k)}\over{2\Gamma^2(-\ell+1)\Gamma(-\ell+2k)}}(-2k)^{\ell+1}M^{-1}\
,
\end{eqnarray}
and
\begin{eqnarray}\label{Eq32}
\psi_2={{2(2\ell+1)\Gamma^2(2\ell+1)\Gamma(2+2k)}\over{\ell(\ell+1)\Gamma^2(\ell)
\Gamma(\ell+1+2k)}}(2k)^{-\ell-2}M \nonumber
\\
-{{2(2\ell+1)(\ell+1)\Gamma^2(-2\ell-1)\Gamma(2+2k)}\over{\ell\Gamma^2(-\ell)\Gamma(-\ell+2k)}}(2k)^{\ell-1}M\
.
\end{eqnarray}

A spatially bounded solution which respects the boundary condition
(\ref{Eq22}) is characterized by $\psi(z\to\infty)\to 0$. The
coefficient $\psi_2$ in (\ref{Eq30}) should therefore vanish, which
yields the resonance condition [see Eq. (\ref{Eq32})]
\begin{eqnarray}\label{Eq33}
(2k)^{2\ell+1}=\Big[{{\Gamma(2\ell+1)\Gamma(-\ell)}\over{(\ell+1)\Gamma(-2\ell-1)\Gamma(\ell)}}\Big]^2
{{\Gamma(-\ell+2k)}\over{\Gamma(\ell+1+2k)}}\ \nonumber
\\
\end{eqnarray}
for the bound-state energies (unstable eigenvalues) of the system.
Substituting into (\ref{Eq33}) the value
$\ell={{-1+i\sqrt{3}}\over{2}}$ for the effective harmonic index
$\ell$ [see Eq. (\ref{Eq13})] and using Eq. 6.1.18 of \cite{Abram},
one can write the resonance condition (\ref{Eq33}) in the form
\begin{eqnarray}\label{Eq34}
k^{i\sqrt{3}}=8^{i\sqrt{3}}e^{i2\pi/3}{{\Gamma^2(i{{\sqrt{3}}\over{2}})}
\over{\Gamma^2(-i{{\sqrt{3}}\over{2}})}}
{{\Gamma({{1-i\sqrt{3}}\over{2}}+2k)}\over{\Gamma({{1+i\sqrt{3}}\over{2}}+2k)}}\
 .
\end{eqnarray}

Since $k$ is a small quantity [see Eqs. (\ref{Eq19}) and
(\ref{Eq23})], one can use an iteration scheme in order to solve the
resonance condition (\ref{Eq34}). The zeroth-order resonance
equation is given by
\begin{eqnarray}\label{Eq35}
k^{i\sqrt{3}}=8^{i\sqrt{3}}e^{i2\pi/3}{{\Gamma^2(i{{\sqrt{3}}\over{2}})}
\over{\Gamma^2(-i{{\sqrt{3}}\over{2}})}}
{{\Gamma({{1-i\sqrt{3}}\over{2}})}\over{\Gamma({{1+i\sqrt{3}}\over{2}})}}\
 .
\end{eqnarray}
Denoting
\begin{equation}\label{Eq36}
\theta=\arg[\Gamma({{i\sqrt{3}}/{2}})]\ \ \ ; \ \ \
\phi=\arg[\Gamma({{1+i\sqrt{3}}\over{2}})]\  ,
\end{equation}
one finds from (\ref{Eq35}) the {\it infinite} set
\begin{eqnarray}\label{Eq37}
M\omega^{(0)}_n=i\times4e^{-{{2\pi}\over{\sqrt{3}}}(n-{1\over3})+{{4\theta-2\phi}\over{\sqrt{3}}}}\
\ ; \ \ n=1,2,3,...\  .
\end{eqnarray}
of zeroth-order unstable ($\Im\omega>0$) eigenvalues \cite{Notenn}.

Substituting (\ref{Eq37}) into the r.h.s of (\ref{Eq34}), one
obtains the first-order resonance condition
\begin{eqnarray}\label{Eq38}
k^{i\sqrt{3}}=8^{i\sqrt{3}}e^{-i2\pi(n-{1\over3})} \nonumber
\\ \times{{\Gamma^2(i{{\sqrt{3}}\over{2}})}
\over{\Gamma^2(-i{{\sqrt{3}}\over{2}})}}
{{\Gamma({{1-i\sqrt{3}}\over{2}}+16e^{-{{2\pi}\over{\sqrt{3}}}(n-{1\over3})+{{4\theta-2\phi}\over{\sqrt{3}}}})}
\over{\Gamma({{1+i\sqrt{3}}\over{2}}+16e^{-{{2\pi}\over{\sqrt{3}}}(n-{1\over3})+{{4\theta-2\phi}\over{\sqrt{3}}}})}}\
 .
\end{eqnarray}
Denoting
\begin{equation}\label{Eq39}
\phi_n=\arg\big[\Gamma({{1+i\sqrt{3}}\over{2}}+16e^{-{{2\pi}\over{\sqrt{3}}}(n-{1\over3})+
{{4\theta-2\phi}\over{\sqrt{3}}}})\big]\
,
\end{equation}
one finds from (\ref{Eq38}) the infinite family
\begin{eqnarray}\label{Eq40}
M\omega^{(1)}_n=i\times4e^{-{{2\pi}\over{\sqrt{3}}}(n-{1\over3})+{{4\theta-2\phi_n}\over{\sqrt{3}}}}\
\ ; \ \ n=1,2,3,...\
\end{eqnarray}
of first-order unstable eigenvalues \cite{Notecu}.

For the first two `excited' eigenvalues one finds from (\ref{Eq40})
\begin{equation}\label{Eq41}
M\omega^{(1){\text{ana}}}_1=1.25\times 10^{-2}i\ \ \ ; \ \ \
M\omega^{{(1)}\text{ana}}_2=3.65\times 10^{-4}i\  .
\end{equation}
For comparison, the corresponding numerically computed eigenvalues
in the large-mass limit (\ref{Eq10}) are given by
\cite{Oliv,Notelarg2}
\begin{equation}\label{Eq42}
M\omega^{\text{num}}_1=1.23\times 10^{-2}i\ \ \ ; \ \ \
M\omega^{\text{num}}_2=3.57\times 10^{-4}i\  .
\end{equation}
One therefore finds a fairly good agreement (to within $\sim 2\%$),
\begin{equation}\label{Eq43}
{{\omega^{{(1)}\text{ana}}_1}\over{\omega^{\text{num}}_1}}\simeq
1.018\ \ \ ; \ \ \
{{\omega^{{(1)}\text{ana}}_2}\over{\omega^{\text{num}}_2}}\simeq
1.023\  ,
\end{equation}
between the analytical formula (\ref{Eq40}) for the unstable
eigenvalues of the black hole and the numerically computed
\cite{Oliv} eigenvalues (\ref{Eq42}) \cite{Notezg}.

\section{Summary and physical implications}

In summary, we have analyzed the instability spectrum of weakly
magnetized
SU(2) Reissner-Nordstr\"om black holes. In
particular, we have derived analytical expressions
for the infinite family of unstable (imaginary) black-hole
resonances.

The main results derived in this paper and their physical
implications are as follows:

(1) For the analysis of the fundamental instability eigenvalue,
$\omega_0$, we have used an expansion method which originally was
developed for the analysis of black-hole quasinormal resonances
\cite{Dol}. Here we have demonstrated that this analytical method
can also be applied successfully to the analysis of fundamental
bound-state energies. The black-hole fundamental instability
eigenvalue $\omega_0$ is given by Eq. (\ref{Eq16}).

(2) For the analysis of the infinitely large spectrum of `excited'
eigenvalues \cite{Noteq}, $\{\omega_n\}_{n=1}^{n=\infty}$, we have
used an appropriate small frequency $M\omega\ll1$ matching procedure
in order to solve the Schr\"odinger-like wave equation (\ref{Eq11})
which governs the dynamics of perturbations in the SU(2)
Reissner-Nordstr\"om black-hole spacetime. The excited instability
spectrum $\{\omega_n\}_{n=1}^{n=\infty}$ is given by the analytical
formula (\ref{Eq40}) \cite{Notezg}.

(3) We have shown that the {\it analytically} derived formulas for
the characteristic instability spectrum of the weakly magnetized
SU(2) Reissner-Nordstr\"om black hole, Eqs. (\ref{Eq16}) and
(\ref{Eq40}), agree with direct {\it numerical} computations
\cite{Oliv} of the eigenvalues.

(4) The interesting numerical work of Rinne \cite{Oliv} has recently
revealed that unstable SU(2) Reissner-Nordstr\"om black holes may
play the role of {\it approximate} \cite{Noteaprr} codimension-two
intermediate attractors (critical solutions) in dynamical
gravitational collapse of the Yang-Mills field. In particular, it
was found \cite{ChCh,ChHi,BizCha,Oliv} that, during a near-critical
gravitational collapse of the Yang-Mills field, the time spent in
the vicinity of the critical solution (that is, the time spent in
the vicinity of an unstable black-hole solution of the
Einstein-Yang-Mills equations) exhibits a critical scaling law [see
Eqs. (\ref{Eq1}) and (\ref{Eq2})], where the critical exponents are
given by the reciprocals of the corresponding instability
eigenvalues.

Our formulas (\ref{Eq16}) and (\ref{Eq40}) provide explicit
analytical expressions for these critical exponents (instability
eigenvalues) in the regime where the weakly magnetized SU(2)
Reissner-Nordstr\"om black hole plays the role of the critical
intermediate attractor \cite{Oliv}. To the best of our knowledge,
this is the first time that a critical exponent of nontrivial
gravitational collapse is calculated analytically.

\bigskip
\noindent
{\bf ACKNOWLEDGMENTS}

This research is supported by the Carmel Science Foundation. I would
like to thank Oliver Rinne for sharing with me his numerical data. I
would also like to thank Yael Oren, Arbel M. Ongo, and Ayelet B.
Lata for helpful discussions.


\begin{thebibliography}{99}
\bibitem{Notehn} These hairy black-hole solutions are labeled by an integer parameter
$n$ which counts the number of nodes of the exterior Yang-Mills
field.
\bibitem{Notepp} Here $|p-p^*|$ is a measure for the distance of the initial data from
the threshold (critical) solution \cite{GunMar}.
\bibitem{Notelms} These are SU(2) Reissner-Nordstr\"om black holes
whose magnetic charges are small on the scale set by the black-hole
mass, see Eq. (\ref{Eq10}) below.
\bibitem{Noteunit} We use natural units in which
$G=c=\hbar=1$.
\bibitem{Notehor} Note that the near-horizon region $r \to r_+$ corresponds to
$x \to -\infty$, whereas spatial infinity $r\to\infty$ corresponds
to $x\to\infty$.
\bibitem{Notewe} Note that $\omega^2$ plays the role of
energy in the Schr\"odinger-like wave equation (\ref{Eq11}). Thus,
eigenfrequencies with $\omega^2<0$ correspond to bound states of the
system.
\bibitem{Notepm} One can also take $\ell\equiv l_{\text{eff}}={{-1-i\sqrt{3}}\over{2}}$
which corresponds to stable modes (with $\Im\omega<0$) that blow up
at $x\to\pm\infty$.
\bibitem{Notebc} Note that the boundary conditions used in \cite{Dol} for the analysis
of the black-hole quasinormal resonances [that is, outgoing waves at
spatial infinity: $\xi(x\to \infty)\sim e^{i\omega x}$, and ingoing
waves at the black-hole horizon: $\xi(x\to -\infty)\sim e^{-i\omega
x}$] correspond in our case [that is, for unstable (bound-state)
modes with $\omega=i|\omega|$] to spatially bounded solutions with
$\xi(x\to \infty)\sim e^{-|\omega| x}\to 0$ and $\xi(x\to
-\infty)\sim e^{|\omega|x}\to 0$.
\bibitem{Noteln} Here we have used the values $\beta\equiv 1-s^2=1-1=0$ ($s^2=1$ corresponds
to electromagnetic perturbation fields) and $N\equiv
n+1/2=0+1/2=1/2$ ($n=0$ corresponds to the fundamental eigenvalue)
in equations (17)-(22) of \cite{Dol}.
\bibitem{Notelarg} This numerically computed \cite{Oliv} eigenvalue corresponds to a
weakly-magnetized SU(2) Reissner-Nordstr\"om black hole with
$r_+=10\gg1$. This horizon radius corresponds to a black-hole mass
of $M=5.05$ [see Eq. (\ref{Eq5})]. These are the largest
horizon-radius and black-hole mass studied numerically in
\cite{Oliv}.
\bibitem{Noteas} Here we have substituted $a=0$ (which corresponds to a non-rotating black
hole), $\lambda=\ell(\ell+1)-s(s+1)$ (which is the angular
eigenvalue corresponding to a non-rotating black hole), and $s=-1$
(which corresponds to electromagnetic perturbation fields) in Eq.
(2.9) of \cite{TeuPre} (known as the master radial Teukolsky
equation). It is worth noting that our final result [see Eq.
(\ref{Eq40}) below] is invariant under the transformation $s\to -s$.
\bibitem{Noteov} Note that the overlap region $1\ll z\ll
1/k$ exists in the small-frequency regime $k\ll1$ [see Eqs.
(\ref{Eq19}) and (\ref{Eq23})].
\bibitem{Notenn} Here we have used the relation $1=e^{-i2\pi n}$ for
$n=1,2,3,...\ $. In addition, we have used here Eq. 6.1.23 of
\cite{Abram}.
\bibitem{Notecu} Curiously, we find the relation
$\theta/2\phi=1.0016$. One can therefore replace, with an accuracy
of $0.05\%$, the term ${{4\theta-2\phi}\over{\sqrt{3}}}$ in
(\ref{Eq37}) and (\ref{Eq39}) by the simpler expression
$\sqrt{3}\theta$. Taking cognizance of the expression (\ref{Eq16})
for the fundamental eigenvalue, which also contains an explicit
factor of $\sqrt{3}$, one may suspect that the relation
${{4\theta-2\phi}\over{\sqrt{3}}}\simeq \sqrt{3}\theta$ (with a
remarkable accuracy of $0.05\%$) may be more than just a mere
coincidence.
\bibitem{Notelarg2} These numerically computed \cite{Oliv} eigenvalues correspond
to a weakly-magnetized SU(2) Reissner-Nordstr\"om black hole with
$r_+=10\gg1$. This horizon radius corresponds to a black-hole mass
of $M=5.05$ [see Eq. (\ref{Eq5})]. These are the largest
horizon-radius and black-hole mass studied numerically in
\cite{Oliv}.
\end{thebibliography}
\end{document}